\documentclass[%
 reprint,
unsortedaddress,
onecolumn,
 amsmath,amssymb,
 aps,
]{revtex4-2}

\usepackage{graphicx}% Include figure files
\usepackage{bm}% bold math
\usepackage{xcolor}
\usepackage{amsthm}
\usepackage{makecell}
\usepackage{siunitx}

\begin{document}

\title{Breaking Reflection Symmetry: Evolving Long Dynamical Cycles in Boolean Systems}

\preprint{APS/123-QED}

\author{Mathieu Ouellet}
\affiliation{Department of Electrical \& Systems Engineering, School of Engineering and Applied Science, University of Pennsylvania, Philadelphia, PA 19104 USA}
\author{Jason Z. Kim} 
\altaffiliation[Present Address: ]{Department of Physics, Cornell University, Ithaca, NY, 14853 USA}
\affiliation{Department of Bioengineering, School of Engineering \& Applied Science, University of Pennsylvania, Philadelphia, PA 19104 USA}
\author{Harmange Guillaume}
\affiliation{Perelman School of Medicine, University of Pennsylvania, Philadelphia, PA, USA}
\affiliation{Cell and Molecular Biology Group, Perelman School of Medicine, University of Pennsylvania, Philadelphia, PA, USA}
\author{Sydney M. Shaffer}
\affiliation{Department of Biological Engineering, School of Engineering \& Applied Science, University of Pennsylvania, Philadelphia, PA 19104 USA}
\affiliation{Cell and Molecular Biology Group, Perelman School of Medicine, University of Pennsylvania, Philadelphia, PA, USA}
\affiliation{Department of Pathology and Laboratory Medicine, Perelman School of Medicine, University of Pennsylvania, Philadelphia, PA, USA}
\author{Lee C. Bassett}
\thanks{These authors contributed equally.}
\email[To whom correspondence should be addressed: ]{lbassett@seas.upenn.edu \& dsb@seas.upenn.edu}
\affiliation{Department of Electrical \& Systems Engineering, School of Engineering and Applied Science, University of Pennsylvania, Philadelphia, PA 19104 USA}
\author{Dani S. Bassett} 
\thanks{These authors contributed equally.}
\email[To whom correspondence should be addressed: ]{lbassett@seas.upenn.edu \& dsb@seas.upenn.edu}
\affiliation{Department of Electrical \& Systems Engineering, School of Engineering and Applied Science, University of Pennsylvania, Philadelphia, PA 19104 USA}
\affiliation{Department of Biological Engineering, School of Engineering \& Applied Science, University of Pennsylvania, Philadelphia, PA 19104 USA}
\affiliation{Department of Physics \& Astronomy, College of Arts \& Sciences, University of Pennsylvania, Philadelphia, PA 19104 USA}
\affiliation{Department of Neurology, Perelman School of Medicine, University of Pennsylvania, Philadelphia, PA 19104 USA}
\affiliation{Department of Psychiatry, Perelman School of Medicine, University of Pennsylvania, Philadelphia, PA 19104 USA}
\affiliation{Santa Fe Institute, Santa Fe, NM 87501 USA}

% From the perfect radial symmetries of radiolarian mineral skeletons to the broken symmetry of homochirality, the logic of Nature's regularities has fascinated scientists for centuries. Some of Nature's symmetries are clearly visible in morphology and physical structure, whereas others are hidden in the network of interactions among system components. Just as visible symmetries and asymmetries contribute to the system's beauty, might hidden symmetries contribute to the system's functional harmony? And if so, how? Here we demonstrate that the interaction networks of Boolean systems display a form of dynamical reflection symmetry that expands the lengths of their dynamical cycles. To probe the conditions under which this reflection symmetry evolves, we use a multi-objective genetic algorithm to produce networks with long dynamical cycles. We find that local structural motifs that break the reflection symmetry are deleted in the evolutionary process, providing evidence for symmetry's causal role in dynamical complexity.

\begin{abstract}
In interacting dynamical systems, specific local interaction rules for system components give rise to diverse and complex global dynamics.
Long dynamical cycles are a key feature of many natural interacting systems, especially in biology.
Examples of dynamical cycles range from circadian rhythms regulating sleep to cell cycles regulating reproductive behavior.
Despite the crucial role of cycles in nature, the properties of network structure that give rise to cycles still need to be better understood.
Here, we use a Boolean interaction network model to study the relationships between network structure and cyclic dynamics.
We identify particular structural motifs that support cycles, and other motifs that suppress them. 
More generally, we show that the presence of \emph{dynamical reflection symmetry} in the interaction network enhances cyclic behavior.
In simulating an artificial evolutionary process, we find that motifs that break reflection symmetry are discarded.
We further show that dynamical reflection symmetries are over-represented in Boolean models of natural biological systems.
Altogether, our results demonstrate a link between symmetry and functionality for interacting dynamical systems, and they provide evidence for symmetry's causal role in evolving dynamical functionality.
\end{abstract}

\maketitle
\newpage

\section*{Introduction}

One of the most intriguing characteristics of complex systems is that they evince emergent global functions from local interactions.
Gene regulatory networks are a quintessential example, describing the complex network of short time-scale interactions between molecules to produce the long time-scale cycles of reactions that sustain life \cite{chen2016boolean,berdahl2009random,iguchi2007boolean,sinha2020behavior}. Such cyclic behaviors play a fundamental role in many processes, including cell cycles \cite{klemm2005topology}, biological clocks \cite{forger2017biological}, cell fate \cite{kobayashi2009cyclic}, cancer regulation and DNA damage \cite{wang2019roles}, and signaling \cite{nash2001receptor}. It is known that cyclic behavior in random Boolean models arises more often than fixed points and that such behavior is favored by evolution \cite{pinho2012most}. Despite their significance and prevalence, long cyclic reactions remain far from understood, in part because the process of determining their underlying mechanisms is made difficult by the nonlinear and heterogeneous distribution of interactions. Can we distill simple principles for how specific patterns of local interactions determine long and complex cycles of reactions?

Biological systems have been fruitfully modeled as Boolean networks to shed light on this question. In these models, the state of each component---a gene, protein, or RNA---is described by a binary value, and the interactions between components---binding, chemical reaction, and so on---are described by Boolean functions. Prior work has extensively studied the interaction functions \cite{aracena2008maximum,mori2017expected,aracena2017number} to model probabilistic \cite{raj2008nature} and multi-level \cite{chen2018novel} interactions or to stabilize existing sequences of reactions \cite{campbell2014stabilization}. Other work has focused on the intensive study of specific network topologies \cite{berdahl2009random, iguchi2007boolean,coppersmith2001reversible, ouma2018topological, drossel2005number,somogyvari2000length} and local structures that are typically referred to as motifs \cite{mangan2003structure}. However, we still lack a general understanding of how the local interaction topology determines long sequences of cycles, thereby limiting our ability to make principled predictions across different networks about the global effects of local structures.

Here, we provide such an understanding through the analytical and numerical study of Boolean network topology. First, we use an evolutionary algorithm to optimize for network motifs with long cycle lengths and discover the existence of \textit{suppressed motifs} that are almost entirely absent in the evolved networks. Next, we discover that many such evolved networks display a \textit{dynamical reflection symmetry}, such that if the network at state $\Vec{x}(t)$ transitions to state $\Vec{y}=\Vec{x}(t+1)$, then that same network at state $\Vec{1}-\Vec{x}(t)$ transitions to state $\Vec{1}-\Vec{y}$. Moreover, we find that the suppressed motifs systematically break this symmetry. 

To demonstrate the practical utility of our finding, we apply it to real biological systems and find that reflection symmetry appears naturally in networks that have  evolved to support long dynamical cycles, whereas suppressed motifs decrement the length of dynamical cycles.  Our findings demonstrate how dynamical symmetries play a crucial role in the observed complexity of biological systems.

\begin{figure*}[h]
\centering
\includegraphics[width=1\linewidth]{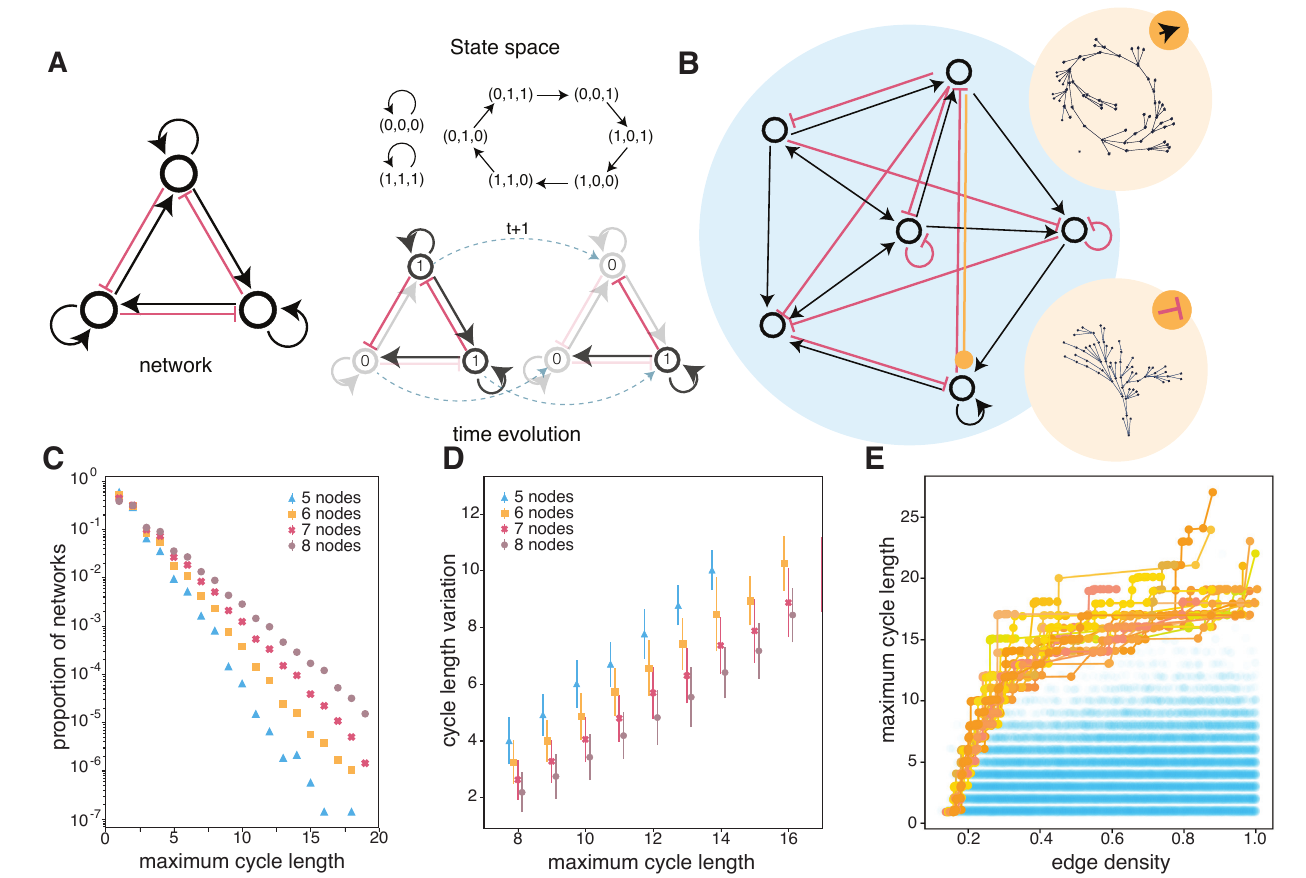}
\caption{\textbf{Boolean network model.} \textbf{(A)}(\emph{left}) Example of an interaction network. Red indicates an inhibitory connection; black indicates an excitatory connection. Curved arrows indicate self-loops. \textbf{(A)}(\emph{bottom, right}) Two consecutive temporal states of the network on the left. Connections not in use are shown in grey. The state of each node is shown as a `0' or `1' inside the relevant circle. \textbf{(A)}(\emph{top, right}) The full state space of the network is shown on the left. Arrows indicate the temporal progression from state to state. Each state is encoded as the activity of the three nodes (e.g., `(0,1,1)' listed clockwise starting from the top one). \textbf{(B)} A larger interaction network of 6 nodes. The state space of this network will depend upon whether the orange edge is excitatory or inhibitory. In the former case, the state space is as shown in the top beige circle; in the latter case, the state space is as shown in the bottom beige circle. \textbf{(C)} Distribution of maximum cycle length for uniformly sampled networks of 5, 6, 7, and 8.
\textbf{(D)} Expected decrement in cycle length ($x$-axis) when a single edge is randomly altered ($y$-axis).  \textbf{(E)} Optimized networks for cycle length obtained by random sampling (blue) and by a Pareto evolutionary algorithm (orange).}
\label{fig:fig1}
\end{figure*}

\section*{Boolean Network Model}

Our Boolean network model follows a typical construction \cite{chen2018novel} motivated by biologically meaningful functions \cite{raeymaekers2002dynamics}, the notion of threshold logic \cite{hu1965threshold}, and multiple models of Boolean networks \cite{greil2007kauffman, wacker2020boolean, pinho2012most} that have been used successfully in biology. 
The system state $\Vec{x}(t)$ is represented by an n-dimensional Boolean vector in $\mathbb{B}^n = \{\mathrm{True, False}\}^n$, a finite space of dimension $2^n$ that we refer as the \textit{state space}. For simplicity, we will often use an integer representation of True as 1 and False as 0. 
We consider networks where interactions between the nodes are either null, inhibitory, or excitatory. We represent those interactions by a weighted adjacency matrix, $\mathbf{A}$ where $A_{ij}$ is $1$ for an excitatory edge from node $i$ to node $j$, $-1$ for an inhibitory one, or 0 if there is no interaction (see Figure \ref{fig:fig1}. B). 

%First, we denote $\mathbb{Z}^n$ as the space containing all $n$-dimensional vectors of integers where the sum $\sum_j A_{ji}x_j(t)$ of equation \ref{eqn:evolvestate} lies. Second, we denote $\mathbb{B}^n = \{\mathrm{True,False}\}^n$ as the space containing the Boolean states, which we later convert to integers by mapping True to 1 and False to 0, thereby forming the state vector $x(t)$. Multiple useful binary and unary operators act on this space such as $\neg, \lor,\land$, the negation, the OR-gate, and the AND-gate, respectively. To determine if the component $x_i(t+1)$ is 0 or 1, we only need to know the sign of the right-hand side of equation \ref{eqn:evolvestate}. All components greater than 0 are mapped to $1$, and the rest are mapped to $0$. 
% To reproduce these mechanisms, we allow edges to be excitatory or inhibitory. Excitatory interactions increase gene expression or protein concentration, whereas inhibitory interactions decrease that expression or concentration. 

The states of all nodes are updated in discrete time steps; all states are updated at once. At every time step, each node sums the excitatory and inhibitory inputs, and if that sum is greater than 0, then the node becomes active (1); otherwise, it becomes inactive (0). The update rule can be formally specified as follows:
\begin{equation}
      x_i(t+1) =
    \begin{cases}
      1, & \mathrm{if} \sum_{j}A_{ji}x_j (t) > 0\\
      0, & \mathrm{if} \sum_{j}A_{ji}x_j (t) \leq 0 
    \end{cases}
    \label{eqn:evolvestate}
\end{equation}
and we will write $\tau(\Vec{x}(t)) =  \Vec{x}(t+1)$.

\section*{Numerical Results}

\subsection*{Sampling networks}
To relate cycling dynamics to the topology of the interaction network, we began by considering random networks. We discovered that most random networks have a short cycle length (Figure \ref{fig:fig1}.C), which is reflected in the exponential tail of the cycle-length distribution. The identification of cycles is an NP-hard problem \cite{bridoux2021complexity}. Hence we limit our study of cycle lengths to networks containing eight or fewer nodes. Long cycles are also extremely rare. For instance, a random network of seven nodes has a cycle of length 19 with an approximate probability of one in a million (see Figure \ref{fig:fig1}.C). We have also found that edge density is a determining factor in maximal cycle length, as denser networks are more likely to exhibit long cycles. Furthermore, altering the nature of a single edge (e.g., from excitatory to inhibitory) has a strong destructive effect on the maximal cycle length (see Figure \ref{fig:fig1}).  We did not observe a correlation between the degree of asymmetry in the connection matrix and the length of cycles; in fact, we found that both the best and worst cycles often exhibit high degrees of symmetry, suggesting that the specific nature of the asymmetry is crucial in determining cycle length.

The apparent simplicity of this Boolean model belies surprising complexity. Unlike Hopfield networks, the lack of symmetry in the interaction matrix ($a_{ij} \neq a_{ji}$) of Boolean networks implies the non-existence of a Lyapunov function, making them difficult to study analytically \cite{hopfield1982neural,hopfield2007hopfield}. Further, the inclusion of self-loops has been shown to increase the number and robustness of attractor states, thereby increasing the complexity of our model's dynamics \cite{montagna2020impact}. For these reasons, such models are remarkably expressive and useful in explaining real biological observations such as cell differentiation \cite{braccini2018self}.

\begin{figure*}[htp]
\centering
\includegraphics[width=0.8\linewidth]{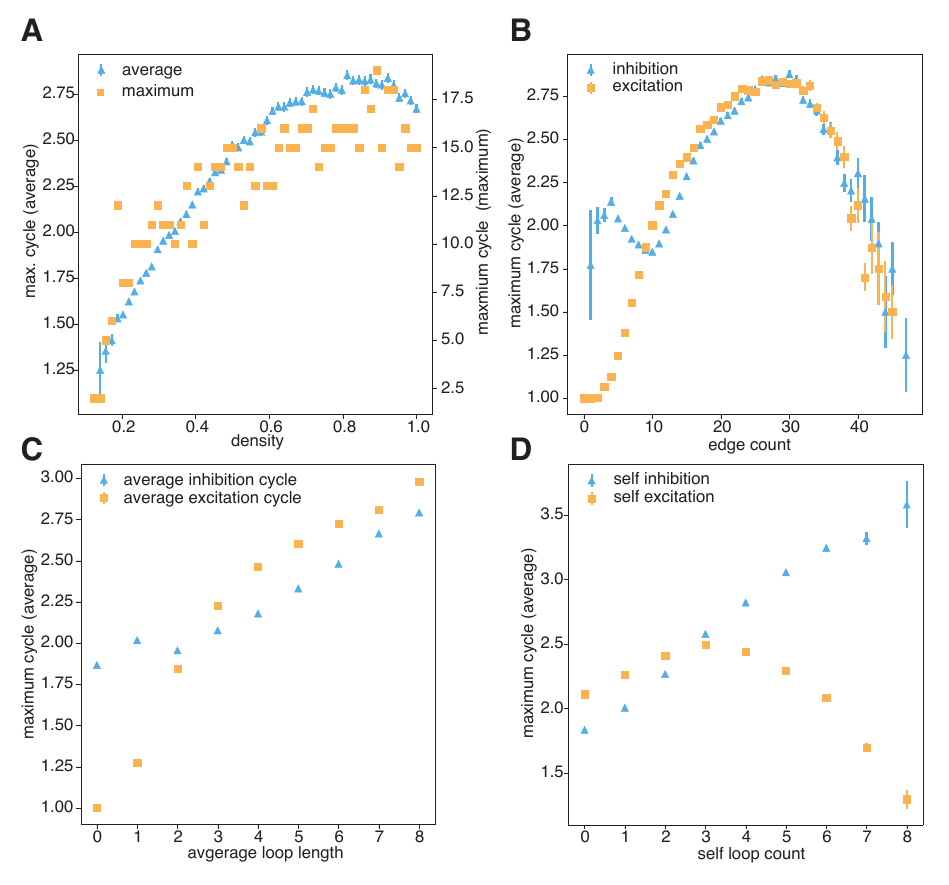}
\caption{\textbf{Topological properties of random networks.} \textbf{(A)} The average and largest maximal cycle length as a function of density for an eight-node network. \textbf{(B)} The average maximal cycle length as a function of the number of edges of each type: excitatory and inhibitory. \textbf{(C)} The average maximal cycle length as a function of the average size of physical cycles within the interaction network. Note that the former is a dynamical property and the latter is an interaction property. \textbf{(D)} The average maximal cycle length as a function of the number of self-loops, as given by the diagonal entries in the adjacency matrix. All plots use a sample of 400,000 random networks.}
\label{fig:suppfig5}
\end{figure*}

We investigated the association between a network's maximum cycle length and its edge density to identify factors influencing cycle length. Our findings indicate that higher edge density leads to an increase in both average and maximum cycle lengths (see Figure \ref{fig:suppfig5}A). We observe that in random networks, a combination of inhibitory and excitatory edges results in longer cycles (see Figure \ref{fig:suppfig5}B). Furthermore, we found that the average maximum cycle length increases with the presence of excitatory or inhibitory circuits in the interaction network (see Figure \ref{fig:suppfig5}.C). We also found that the specific distribution of excitatory and inhibitory connections, particularly in self-connections, differentially affected cycle length. Self-inhibition, which we defined as an inhibition circuit of length 1, was positively correlated with a high maximum cycle length. By contrast, self-excitation, which we defined as an excitatory circuit of length 1, was negatively correlated with a high maximum cycle length (see Figure  \ref{fig:suppfig5}.D).

To identify and study highly cyclic networks, we utilized an objective function that balances the maximization of cycle length and the minimization of network density. This trade-off is of significance in various domains. Biological systems often face energetic constraints on interactions or the physical pathways on which they rely \cite{bullmore2012economy,gang2016metabolic}, while the boundaries of their design limit engineered systems. The goal of minimizing interactions is countered by the goal of facilitating a diverse dynamical repertoire \cite{zanudo2013effective,egger2020local,jaeger2012life,yamamoto2018impact}. 

We used a genetic Pareto algorithm \cite{deb2002fast} that encodes each network's genetic representation as a string of $n^2$ characters in the set ${-1,0,1}$ (see also Supp. Figure \ref{fig:suppfig6}. A-B-C ). This algorithm finds Pareto efficient solutions without the need to define one objective function encapsulating the two objectives, allowing us to analyze the entire landscape of optimal solutions. Using this genetic algorithm, we evolved random networks along the Pareto front and confirmed that we produced networks with larger maximum cycle lengths than equi-dense random networks (Figure \ref{fig:fig1}. E; for further information regarding the structure of Pareto networks, see Supp. Figure \ref{fig:suppfig6}. D-E).

\begin{figure}[h]
\centering
\includegraphics[width=0.99\linewidth]{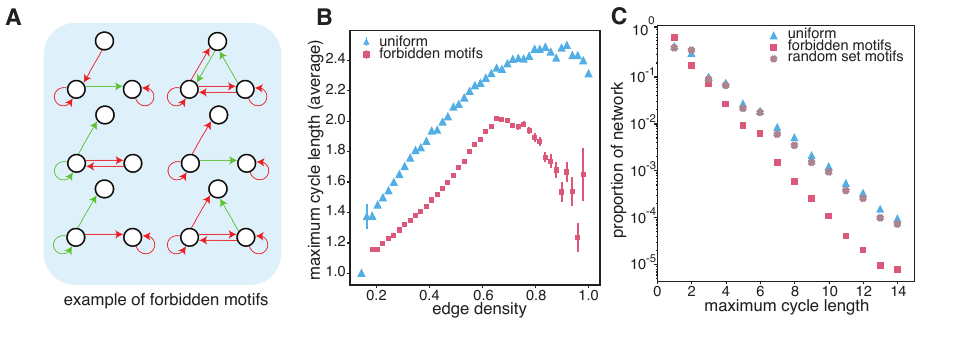}
\caption{\textbf{The role of suppressed motifs in cycle dynamics.} \textbf{(A)} Example of suppressed motifs. For the full list, see Appendix \ref{app:sampling}.  \textbf{(B)} Average cycle length in randomly sampled networks for $n=7$ from a uniform distribution over the space of all interaction network topologies (blue) and from a random sample over networks created by gluing suppressed motifs together (see Appendix \ref{app:sampling}). \textbf{(C)} Proportion of network with a given cycle length for random sample over networks created by gluing together a randomly selected subset of all motifs (red) and for a random subset of motifs. Here we see that the decrease in cycle length is not caused by the gluing process but by the motifs themselves.}
\label{fig:fig2}
\end{figure}

\subsection*{Suppressed motifs} 

We then investigated the factors contributing to the variation in the set of Pareto optimal networks compared to the general population of networks. We found significant differences in the networks' local topology, which differs markedly between evolved and random networks. We observed that the Pareto front networks' global properties---specifically, their density and average degree---were similar to those of random networks. Yet, their local topology was dramatically different. When considering 3-motifs, i.e., subsets of three nodes in the graph with their connections, we discovered a subgroup of around thirty (out of a possible 3284) 3-motifs that were almost completely absent in the evolved networks (see Figure \ref{fig:fig2}.A and Supp. Figure \ref{fig:motiflist_fig}.A). The existence of these \emph{suppressed motifs} suggests a condition on the network's local connectivity that affects its evolved functionality.

% We sought to identify the mechanism underlying long cycling behavior.

To evaluate the impact of suppressed motifs on cycle length, we artificially created networks containing a high density of suppressed motifs (see the ). We found that the average cycle length was significantly lower than expected in random networks for all network densities ($p<0.0001$) (Figure \ref{fig:fig2}.B). We also found that the density of networks with a given maximum cycle length was lower than expected in random networks (Figure \ref{fig:fig2}. C). To determine the specificity of the observed behavior, we next constructed networks containing a high density of randomly selected 3-node motifs. In this new population, we observed only a 20 \% decrease in the density of networks with a long cycle length; this is in comparison to a more than 95\% decrease for suppressed motifs networks (Figure \ref{fig:fig2}. C; see Supp. Figure \ref{fig:pval_fig} for $p$-values and confidence intervals). These findings suggest that a decremented cycle length is specific to suppressed motifs and is not purely an artifact of sampling over a limited number of motifs.

\begin{figure}[h]
\centering
\includegraphics[width=0.7\linewidth]{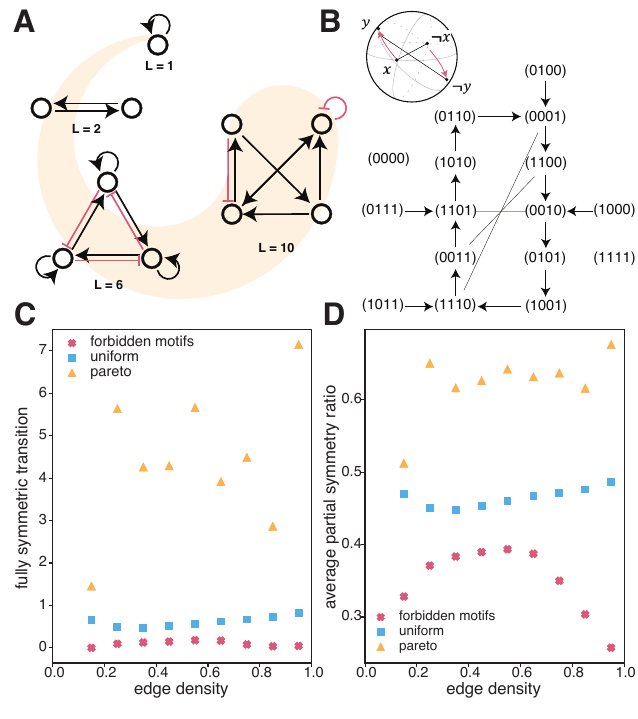}
\caption{\textbf{The role of symmetries in cycle dynamics.} \textbf{(A)} Example networks of up to 4 nodes, with the highest cycle length possible in that network size.  \textbf{(B)} The state space of the $n=4$ network shown in panel (A). The lines without arrowheads represent the states linked under the reflection symmetry (e.g., `(0,0,0,1)' is linked to `(1,1,1,0)'). The top right schematic shows conceptually how the reflection symmetry affects the system's dynamics where a state $x$ is mapped under time evolution to the state $y$. \textbf{(C)} Average number of fully symmetric transitions. \textbf{(D)} We sample the average partial symmetry ratio, the fraction of bits that transition symmetrically, for random networks (blue), evolved networks (orange), and suppressed motif networks (red). }
\label{fig:fig3}
\end{figure}

\subsection*{Dynamical reflection symmetry drives long cycles} 

To better understand what drives the existence of suppressed motifs, we exhaustively enumerated all networks for $n<5$ (Figure \ref{fig:fig3}. A). This becomes rapidly pointless for larger $n$ since the number of networks is approximately given by $\frac{3^{n^2}}{n!}$. For each $n<5$, we identified the networks with the maximal cycle length (see Appendix \ref{app:sampling}). What do all of these networks have in common? We might naively posit that symmetry in the structure of the interaction is important, and indeed the optimized networks with $1<n<4$ exhibit structural symmetry. However, we observed that the 4-node networks with the maximum cycle length were not structurally symmetric, motivating the need for a different explanation. 

As an alternative, we considered a dynamical reflection symmetry that manifests in the network's state-space representation. Such a reflection symmetry permits the inverse of a sequence of states as another sequence.
The inverse state is given by the $\neg$ operator or the standard NOT gate. Under this operator, the state of four Boolean nodes $ \Vec{x}=(0110)$ becomes the state $\neg \Vec{x} = \Vec{1}-\Vec{x} = (1001)$. Then, when we say \textit{dynamical} reflection symmetry, we mean that if the system's dynamics evolve to map $x(t)$ to $x(t+1)$, then the system's dynamics also evolve to map $\neg x(t)$ to $\neg x(t+1)$ as follows:
\begin{equation}
    \tau(x) = y \iff \tau(\neg x) = \neg y. 
    \label{eqn:mirror_sym_main_text}
\end{equation}
We observed dynamical reflection symmetry in the network's state transition diagram in all of the $1<n<5$ networks found to have maximal cycle lengths (Figure \ref{fig:fig3}.A-B). 

This observation motivates the question: Might reflection symmetry relate to suppressed motifs, and if so how? We found that the maximum cycle length was significantly lower (Figure \ref{fig:fig3}. C) in the suppressed motif networks than in the random networks (pairwise two-sided $z$-test, $p<0.0001$). In contrast, our evolved networks---built to optimize the maximum cycle length---displayed a 3- to 7-fold increase in reflection symmetric transitions compared to random networks. These findings link reflection symmetry with the presence of suppressed motifs and a network's ability to display cyclic behaviors. 

To better understand how reflection symmetry might relate to suppressed motifs, we returned to our artificially constructed networks containing a high density of suppressed motifs and measured the number of reflection-symmetric transitions. The number of perfectly symmetric transitions is small; hence, it is of value to also estimate partial symmetries. Specifically, we estimate the fraction of bits that respects symmetry. Specifically, given a Boolean network of size $n$, the partial symmetry ratio $p_{sr}$ for this network is given by:
\begin{equation*}
    p_{sr} =  \frac{1}{2^n\cdot n}\sum_x \sum_i \theta(  \lvert f(x)_i - f(\neg x)_i  \rvert ),
\end{equation*}
where $\theta(x)$ takes the value 1 if $x>0$ and is zero elsewhere. With this broader definition of partial symmetry, we observed a similar trend in which evolved networks showed an increase in the partial symmetry ratio, whereas suppressed motif networks showed a decrease (Figure \ref{fig:fig3}.D). These findings motivate further investigations into the causes that might drive correlations between dynamical reflection symmetry, suppressed motifs, and long cycles.

\begin{figure}[h]
\center
\includegraphics[width=0.9\linewidth]{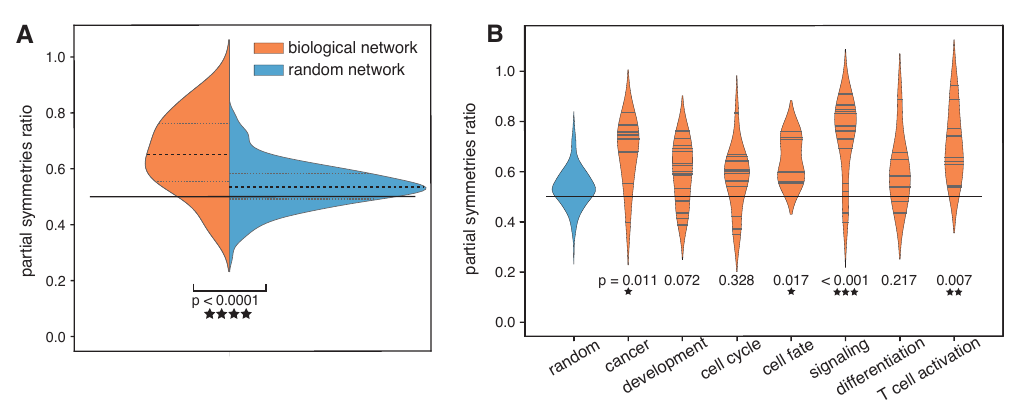}
\caption{\textbf{Reflection symmetry in gene regulation networks.} \textbf{(A)} Comparison between the average symmetry ratio for Boolean network models of biological systems and their random counterparts built to maintain the joint distribution of node number and nodes' in-degree. \textbf{(B)} Comparison between the average symmetry ratio of Boolean network models of biological systems separated into categories according to tags in the GINim repository.}
\label{fig:fig4}
\end{figure}

\subsection*{Real boolean biological networks support reflection symmetries}

Evolved networks that displayed a long maximum cycle length tended to express reflection-symmetric transitions more frequently than random networks. Interestingly, those transitions were not exclusively present inside the cycles but were also found in the non-cyclic part of the state space. This observation suggests that dynamical symmetries may play an even more profound role in evolved networks, and motivates investigation of their presence in broader categories of dynamical networks, both synthetic and natural. We turned to gene regulatory networks to test our hypothesis regarding the presence of reflection symmetries in biology. We used two repositories: the GINsim software \cite{naldi2018logical} and the PyBoolNet python package \cite{klarner2017pyboolnet}. 

The set of Boolean biological networks used contains 70 networks. Networks containing non-binary states were transformed to a binary representation using GINSIM. They have between 3 and 218 nodes, with an average of 33.7 nodes. The average connectivity is in the range  of $[1.18, 4.82]$ with an average of 2.55 connections. The 129 random Boolean networks were generated using the BoolNet package with uniform function generation, and uniform linkage based on Kauffman’s method \cite{kauffman1969metabolic}. The number of genes was selected to reproduce the distribution found in the real network, and the function generation was done randomly using the homogeneous policy. The average number of inputs was selected to reproduce the average number found in the real networks with the given number of nodes.

Using these data, we observed that Boolean models of biological systems showed markedly more symmetries than random networks (Figure \ref{fig:fig4}. A). Specifically, the average number of symmetric transitions in the biological model networks was 64.4\%, whereas the average number in random networks was 53.6\%. The slight deviation from 50\% is due to the finite size of the analyzed networks. Further, the reflection symmetry ratio was significantly greater in the biological networks than in the random networks (two-sided $t$-test, $t=7.068$, $p= \num{4.6e-10}$). In addition to this overall effect, we noted a marked variation across the different models: for 10 of the biological networks, less than 50\% of transitions were symmetric; for 29 of them, more than 70\% were symmetric; and for 12 of them, more than 80\% were symmetric. To better understand this variability, we divided the models into biologically relevant categories using the tags provided in the GINsim repository. The seven most populated categories were retained for analysis (Figure \ref{fig:fig4}. B). We observed that the most symmetrical categories were cell signaling, cell fate, cell activation, and cancer, with only a few networks lying under the 50\% line. 

By considering symmetries in Boolean networks, we illustrate a simple method to determine reflection symmetry and examine the overall development of the system. Notably, this concept of reflection symmetry can also be applied to non-binary state systems and even to systems where only part of the system's evolution is known. As an example, we have included a basic case study using continuously valued scRNA-seq data of cancer cells, demonstrating how these symmetries can be utilized to classify cells that respond to drugs versus those that are drug resistant (see Appendix \ref{app:data}).

\section*{Discussion}
     
\subsection*{Functional Role for Reflection Symmetry.} 
In this study, we uncovered a new link between the expression of specific motifs and the existence of cycling behavior in Boolean networks. Potential links between these two properties have been reported previously for some gene regulatory networks \cite{burda2011motifs}. For example, prior studies report a link between bi-fan motifs and cycling behavior \cite{burda2011motifs} and find that chaotic motifs are linked to cycling \cite{zhang2012chaotic}. Building upon these observations, we have shown that abnormally underrepresented motifs have a specific function as reflection-symmetry breakers. This relationship sheds light on how an interaction network with these motifs can maintain long cycles. 
 
\subsection*{Diversity of Dynamical Symmetries.} 
Here we investigated the problem of dynamical symmetries in Boolean networks. Many other types of symmetries exist. For example, one might seek to understand symmetries in the output function on each node and consider that the function $f_1(\vec{x}) = (x_1 \text{ or } x_2)$ is invariant under permutation of its argument. In fact, output function symmetries at the node level are a powerful way to characterize complex network dynamics \cite{hossein2014symmetry, stearns2019symmetry}. Our work extends these observations by showing that symmetries at the global level can explain some properties of networks with complex dynamics. Another common approach is the study of symmetric properties inside the interaction network, such as fibration symmetry \cite{morone2020fibration, leifer2020circuits}. Our approach differs from these studies in evaluating symmetry in the state space \cite{cantor2014towards}, but nevertheless provides insights regarding a specific property of the interaction network. This property does not appear as a symmetry in the structure but as a balancing equation on each node. By taking a different perspective from prior studies, our approach sheds light on new mechanisms of dynamical symmetries and the function of the systems that support them. 

\subsection*{Methodological Considerations.} 
Several methodological considerations are pertinent to our work. First, there remains a conceptual and formal distinction between a system and its Boolean network model. Here, we have shown that biologically inspired Boolean networks display a high level of reflection symmetry and that not all biological processes have the same amount of symmetry. Yet, it remains unknown whether reflection symmetry is intrinsic to the system or a result of the map between a set of experiments and its Boolean network model. Future theoretical work could examine the impact of the mapping process on our findings. Further experimental work could examine the evolution of a simple biological system (e.g., yeast) and confirm the existence of reflection-symmetric states. 

Boolean models of biological systems can include more general types of interactions and update rules than the ones we considered. Reflection symmetry can be easily identified in both thresholded and un-thresholded models. Furthermore, since every Boolean model maps binary states to binary states, our reflection symmetry definition is broadly applicable and does not depend on the nature of the updating scheme. The difference in models does affect the ability to analyze motifs since the model determines the possible types of interactions. Further work should examine this potential dependence and the broader impact of the updating schemes and the model on the condition necessary for mirror symmetry. Future work could also seek to generalize and test the theory of dynamical reflection symmetry for more general Boolean networks and multiple updating schemes and to random Boolean networks \cite{drossel2005number,liu2006emergent}. Finally, various other deterministic  network-based systems exhibit cyclical behavior, and it would be intriguing to investigate whether subclasses of motifs also exist in systems such as coupled maps on networks \cite{kaneko1992overview, killingback2013competitively} or evolutionary games on networks \cite{killingback1998self}.

\section*{Conclusion}

In identifying the important role of dynamical symmetry in Boolean networks, this work suggests strategies that can be used to engineer complex dynamical systems with particular dynamical features, or to modify an existing system's architecture to influence its properties. 
In particular, we have shown is is possible to enhance or suppress cycling behaviors by altering only a small subset of edges in a system's interaction network topology. 
As applied to complex biophysical systems in pharmacology and microbiology, this understanding may aid the design of targeted diagnostics and therapeutics.

Our findings can also improve Boolean network modeling of real systems.
In situations where cycling behavior is a defining system characteristic, the search for a suitable Boolean network model can be dramatically simplified by considering the symmetric subspace. 
This is crucial since it is impractical to explore the complete state space experimentally, even for small networks. 
Similarly, reflection-symmetric counterparts of observed trajectories can be added to training data sets, doubling the information available to machine learning models at no extra cost, analogously to mirroring images for neural network training.

In elucidating the formal relationship between reflection symmetry and cyclic behaviors, our work raises intriguing new questions for the scientific community. For example: What causes and regulates dynamical symmetries? Why does the degree of dynamical symmetry vary across different cellular and molecular processes? Which diseases or other perturbations to the system impact this symmetry? Is symmetry explained by evolution or caused by the physics of the system? Answers to these questions will profoundly impact our fundamental understanding of natural dynamical systems.

\subsection*{Data and code availability}
All data and code used in the manuscript are available at https://github.com/ouelletmathieu/BMS.

\section*{Acknowledgement}
M.O. acknowledges the support of the Natural Sciences and Engineering Research Council of Canada (NSERC). J.Z.K. acknowledges support from the NIH T32-EB020087. L.C.B. and M.O. acknowledge support from the National Science Foundation through the University of Pennsylvania Materials Research Science and Engineering Center (MRSEC) (DMR-1720530).
D.S.B. acknowledges support from the National Science Foundation (IIS-1926757, DMR-1420530), the Paul G. Allen Family Foundation, and the Army Research Office (W911NF-16-1-0474, W011MF-191-244). The content is solely the responsibility of the authors and does not necessarily represent the official views of any of the funding agencies.

\section*{Citation Diversity statement}
Recent work in several fields of science---including physics and biology, where our work here is situated \cite{chatterjee2021gender, fulvio2021imbalance,dworkin2020extent,teich2021citation}---has identified a bias in citation practices such that papers from women and other minority scholars are under-cited relative to the number of such papers in the field \cite{wang2021gendered,dion2018gendered,mitchell2013gendered,maliniak2013gender,caplar2017quantitative}. Here we sought to proactively consider choosing references that reflect the diversity of the field in thought, form of contribution, gender, race, ethnicity, and other factors. First, we obtained the predicted gender of the first and last author of each reference by using databases that store the probability of a first name being carried by a woman \cite{dworkin2020extent,zhou_dale_2020_3672110}. By this measure (and excluding self-citations to the first and last authors of our current paper), our references contain 5.64\% woman(first)/woman(last), 13.33\% man/woman, 11.4\% woman/man, and 69.64\% man/man. This method is limited in that a) names, pronouns, and social media profiles used to construct the databases may not, in every case, be indicative of gender identity and b) it cannot account for intersex, non-binary, or transgender people. Second, we obtained predicted racial/ethnic category of the first and last author of each reference by databases that store the probability of a first and last name being carried by an author of color \cite{sood2018predicting}. By this measure (and excluding self-citations), our references contain 29.6\% author of color (first)/author of color(last), 11.7\% white author/author of color, 16.78\% author of color/white author, and 41.93\% white author/white author. This method is limited in that a) names and Florida Voter Data to make the predictions may not be indicative of racial/ethnic identity, and b) it cannot account for Indigenous and mixed-race authors or those who may face differential biases due to the ambiguous racialization or ethnicization of their names. We look forward to future work that could help us to better understand how to support equitable practices in science.

\newpage
\clearpage
\newpage
\appendix

\section{Sampling}
\label{app:sampling}
We sample the space of Boolean networks in two steps. First, a density is chosen uniformly at random from the range 0.2 to 1. Second, we select elements in the matrix and fill those elements with either a $-1$ or $+1$---with equal probability---until we reach the target density. For the biased case, we again begin by choosing a  density randomly from 0.2 to 1. Then, we select valid motifs randomly from the set of desired motifs and place them randomly in the interaction network. If we cannot find a position for a given motif, then we discard that motif and randomly draw another from the same set. Once the desired ratio of motifs is achieved, the network is filled with random edges to achieve the global density. Edges used for a motif (empty edges included) are protected and cannot be filled in this part of the process.

\begin{figure*}[htp]
\centering
\includegraphics[width=0.95\linewidth]{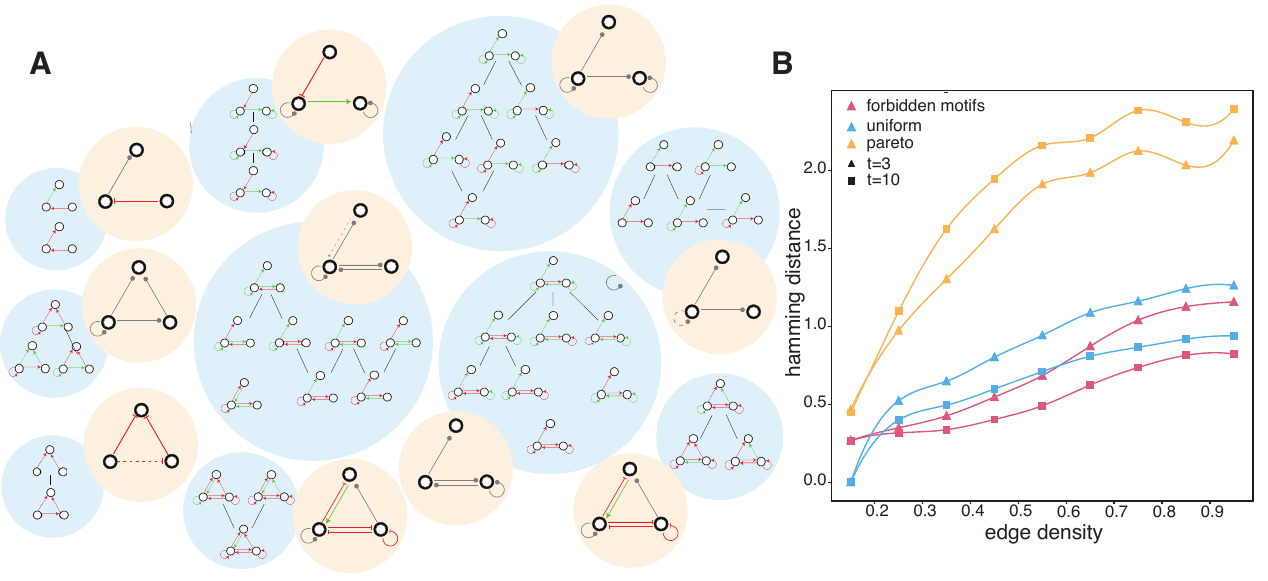}
\caption{\textbf{Ruppressed motifs and their impact on the propagation of perturbations.} \textbf{(A)} Set of suppressed motifs. To improve the readability of this visualization, the set of suppressed motifs was approximately clustered by structural similarity. Each blue circle represents a cluster. The line between motifs inside a cluster shows how motifs in the cluster are related. The beige circle represents the global structure of the motifs inside the blue cluster. \textbf{(B)} Small perturbations propagate more in evolved networks than in random and suppressed motif networks. We sample the average distance between two states which initially differs only by one component after 3 updates (triangle) and after 10 updates (square).} %this is fig 7
\label{fig:motiflist_fig}
\end{figure*}

\subsection*{Evolved networks}

In our evolutionary algorithm, we set the population size to 600 interaction networks with a density taken from a uniform distribution ranging from 0.2 to 1.0. The selection for mating is handled by the NSGA2 algorithm \cite{deb2002fast}. In each step, 400 new offspring are created using two mating operators (see next section for details). We mutate each edge of the offspring's interaction networks with a probability of $P=0.02$. When a component is selected for mutation, its value is changed to one of the two other values in $\{-1,0,1 \}$. The generation process is then repeated until convergence. The whole process is repeated 30 times. 
  
We use two mating operators that reflect the exploratory and convergent mating strategies, respectively. Both operators are used with equal probability each time we call the mating procedure. The first mating operator generates two offspring. For each index of the matrix, each offspring gets assigned either the first or second parent's component. If the first offspring receives its component from parent two, then the second offspring receives its component from parent one. Then both matrices are tested for validity; if one is found to be disconnected when considering the interaction matrix as an undirected graph, the process is restarted. The second mating operator generates one offspring at a time. We list the cycle basis of both parents' interaction networks. Then, random cycle structures are picked from the cycle basis to build a new interaction network containing a random mixture of cycles from both parents. Only cycles that agree for all of their common inhibition or excitation edges are selected. If two selected cycles cannot be joined because of disagreeing edges, the process is restarted. We use the first mating operator if the process fails more than 1000 times. Such failures happen but are rare and represent a negligible percentage of the offspring production. 

\begin{figure*}[htp]
\centering
\includegraphics[width=1\linewidth]{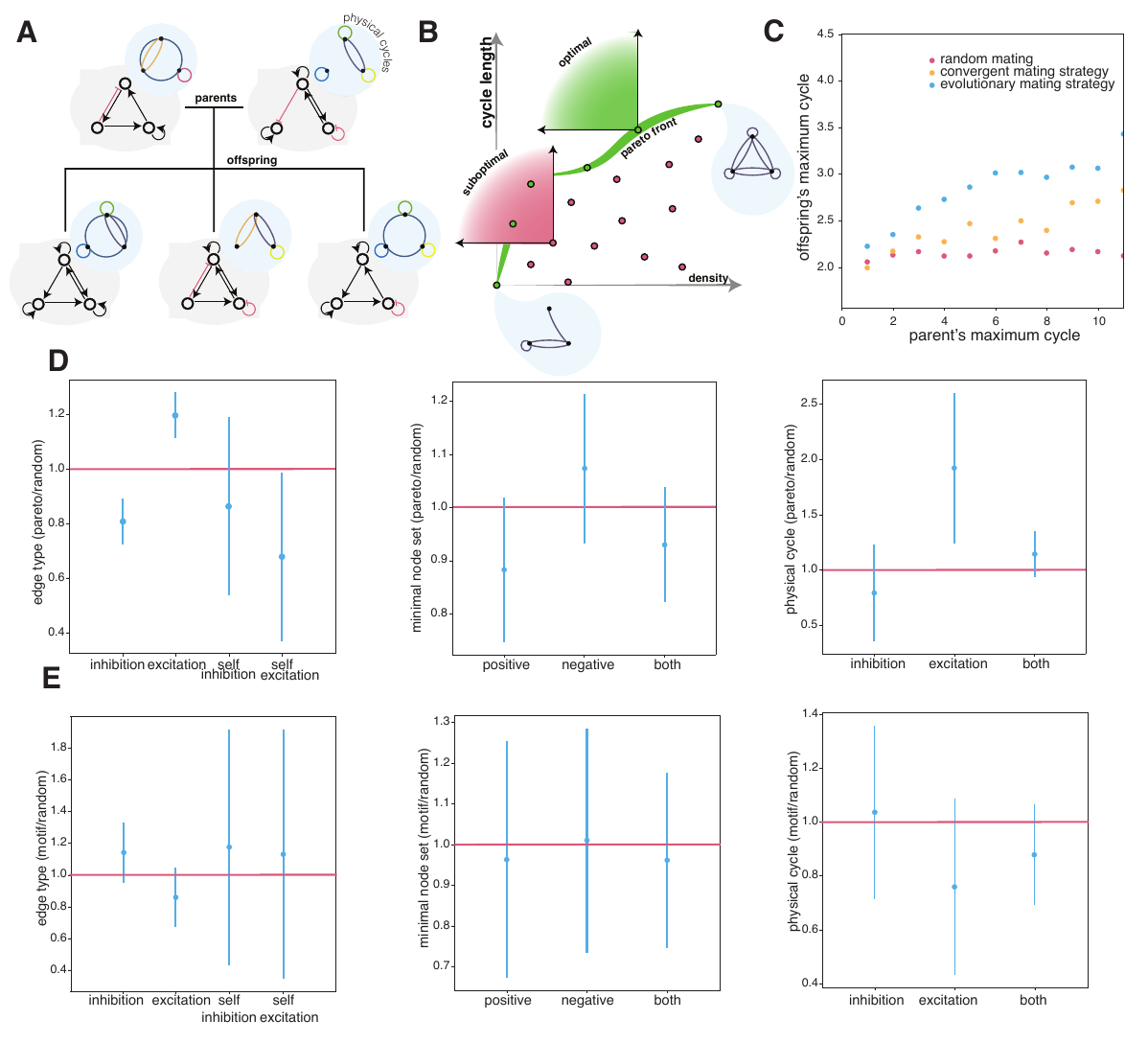}
\caption{\textbf{Offspring generation and properties of suppressed motifs.} \textbf{(A)} Offspring generation is based on cycle preservation. \textbf{(B)} Example of the approximated Pareto front where green points are optimal in the Pareto sense, and where red points are sub-optimal. \textbf{(C)} Evaluation of the different mating strategies or policies that generate offspring. \textbf{(D)} Ratio of the number of different physical constituents of the Pareto front networks versus random networks. \textbf{(E)} Ratio of the number of different physical constituents for the suppressed motif networks versus the random networks.}%this is fig 9
\label{fig:suppfig6}
\end{figure*}

Interestingly, each node in the evolved networks had approximately 20\% more incoming excitatory edges than inhibitory edges. The evolved networks also exhibited nearly twice as many physical excitation circuits as the random networks (see Supp. Figure \ref{fig:suppfig6}.D). These features were emergent properties, as the evolutionary algorithm did not explicitly optimize for them. In addition to comparing all evolved networks to all random networks, we separately examined the dynamics of evolved networks that implemented the exploratory versus the convergent mating strategy.  We found that convergent mating consistently produced networks with greater maximum cycle length than exploratory mating (see Suppl. Figure \ref{fig:suppfig6}.C). In general, the structure of the two parents does not generate offspring with long cyclic behavior. Even for the convergent mating strategy, most cycles from both parents do not act cooperatively, generating important attractors leading to fixed points. 

\subsection*{Motif characterization}
To calculate the $z$-value for the number of expected motifs, we used sampled data to obtain each motif's average expected number and its expected standard deviation. Since the expected number for 3-motifs is always one order of magnitude less than the standard deviation, we consider motifs with an individual $z$-score of less than $-0.1$ to be suppressed. This repression typically corresponds to the absence of the motif in the graph. We then consider a given motif to be suppressed in the population if it is tagged as suppressed in at least 50\% of the population's individual graphs.

\subsection*{Reflection symmetry}
Since for each $n\leq 4$, there was at least one network with the maximal possible cycle length with this symmetry, it is quite tempting to conjecture that there always exists a maximal network with this property. This, of course, cannot be confirmed computationally and could only be realistically verified up to $n=5$ without developing a more sophisticated search algorithm.

\subsection*{Robustness}
\label{supp:robust}
Another frequently cited property of the evolved networks is their robustness to perturbation. We sought to determine if the evolved networks and the suppressed motif networks had distinct levels of robustness. The notion of robustness depends upon a notion of response to perturbation. The distance between the non-perturbed and perturbed states is measured at each time step by the Hamming distance, which only counts the number of non-matching bits. A robust network is one where the distance stays constant or decreases over time. We expect evolved networks to show an increase in robustness, and we expect suppressed motif networks to show a decrease in robustness. The results are surprising: for a large range of edge densities, perturbations to suppressed motif networks do not drive a large divergence in state dynamics in comparison to uniform random networks (Supp. Figure  \ref{fig:motiflist_fig}).

\begin{figure}[htp]
\centering
\includegraphics[width=0.75\linewidth]{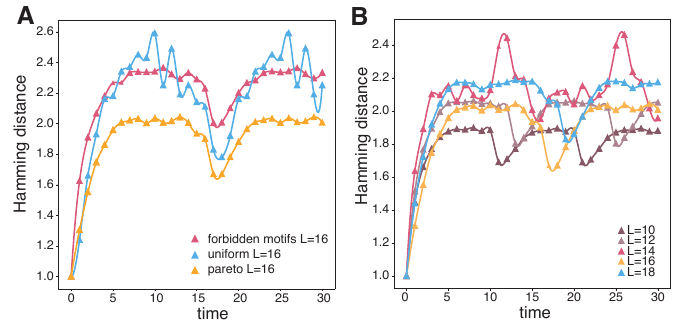}
\caption{\textbf{Time evolution of perturbations.} \textbf{(A)} Average Hamming distance from the 1-bit perturbed network to the unperturbed network as a function of the time for the Pareto front networks binned by cycle length. \textbf{(B)} Same as in panel (A) but for a maximal cycle length of 16 and plotted separately for Pareto front networks, random networks, and suppressed motif networks.} %this is fig 10
\label{fig:suppfig7}
\end{figure}

Intuitively, one might imagine that a network's sensitivity to perturbation could depend upon the maximum cycle length, such that the longer the maximum cycle the greater the tendency for perturbations to drive divergent dynamics. In contrast to this intuition, however, random and suppressed motif networks have quite different average cycle lengths. The difference could then be entirely caused by the presence of long cycles in the evolved population in comparison with the two other populations. To test this possibility, we have compared networks from each category with the same cycle length. For a given maximum cycle length, evolved networks have a slightly lower distance than the random and the suppressed networks for a given cycle length (see Supp. Figure \ref{fig:suppfig7}.A). The difference however is quite small and is cyclic with the period equal to the cycle length, as expected (see Supp. Figure \ref{fig:suppfig7}.B). Therefore, suppressed motifs and the optimization process have, for a given maximum cycle length, a rather low impact on how the networks handle perturbations.

\begin{figure}[htp]
\centering
\includegraphics[width=0.75\linewidth]{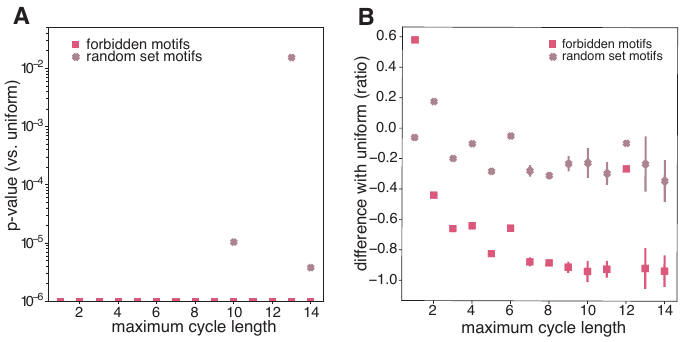}
\caption{\textbf{Statistical analysis.} \textbf{(A)} The $p$-values for the comparison with the uniformly sampled networks shown in Figure \ref{fig:fig2}.D. The $p$-values were capped at $10^{-6}$ when their values were smaller than this number. \textbf{(B)} Proportion difference between the uniformly sampled networks and the motifs enriched networks with the confidence interval shown as error bars.} %this is fig 8
\label{fig:pval_fig}
\end{figure}

\section{Deriving reflection symmetry directly from data} 
\label{app:data}
Boolean networks permit a simple, general definition of dynamical symmetry. However, Boolean models are not available for many  natural dynamical systems, and creating them requires extensive work. Hence, here we analyze symmetry directly from data, circumventing the challenges of Boolean model construction. 

We considered the problem of drug resistance in melanoma at the single-cell level. The state of gene expression can predict which cells will resist therapy and which ones will be sensitive and responsive to therapy. Notably, this prediction is possible before treating the cells with any drug. Thus, we define the state of a cell as a ``primed'' resistant state if that cell would be resistant upon the application of a drug \cite{shaffer2017rare}. For this experiment, we profiled the gene expression of these cells using scRNA-seq. We then divided the cells into two categories: the drug-sensitive cells and the cells primed for drug resistance \cite{shaffer2017rare}. 

\begin{figure}[htp]
\centering
\includegraphics[width=0.6\linewidth]{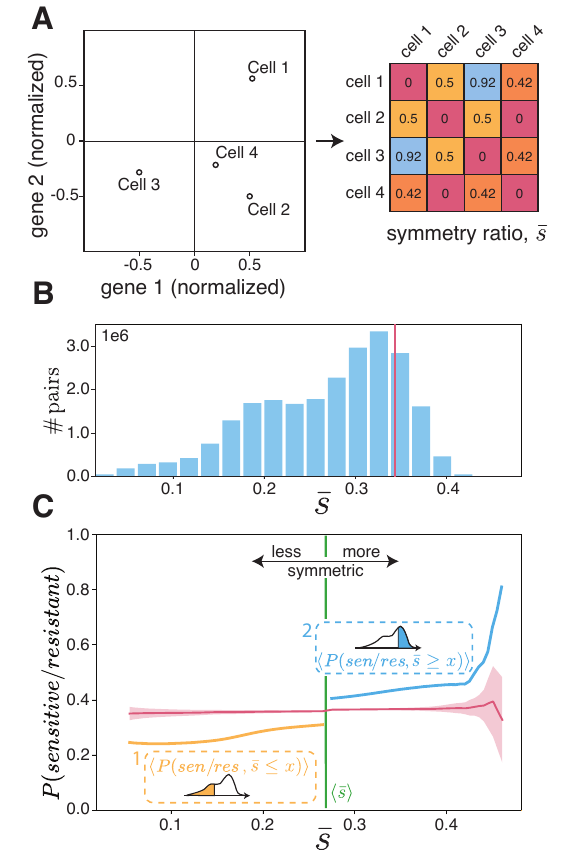}
\caption{\textbf{Reflection symmetry in cancerous cells.} \textbf{(A)} An illustrative example of the reflection symmetry ratio for four cells that express two genes. (\emph{Left}) The normalized level of expression for all four cells and both genes. (\emph{Right}) A symmetric matrix showing the computed reflection symmetry ratio between every pair of cells. In this example, cells 1 and 3 are highly reflection-symmetric, whereas cells 2 and 4 are not reflection-symmetric (asymmetric). \textbf{(B)} The distribution of the reflection symmetry ratio ($\bar{s}$) for all pair of cells. The red line indicates the expectation for uncorrelated, normally distributed genes. \textbf{(C)} The probability for a randomly selected pair of cells to have different states (drug-sensitive state/drug-resistant state) as a function of the reflection symmetry ratio $\bar{s}$. The orange curve shows that the probability for pair of cells with a reflection symmetry ratio $\bar{s}$ less than $x$ as shown in Inset 1. The blue curve shows the probability for pair of cells with a reflection symmetry ratio  $\bar{s}$ higher than $x$ as shown in Inset 2. The red curve shows the probability obtained by random permutations of the labels for both cases. The green line shows the average reflection symmetry ratio over all the pairs of cells. }
\label{fig:fig5}
\end{figure}

\begin{figure}[htp]
\centering
\includegraphics[width=0.6\linewidth]{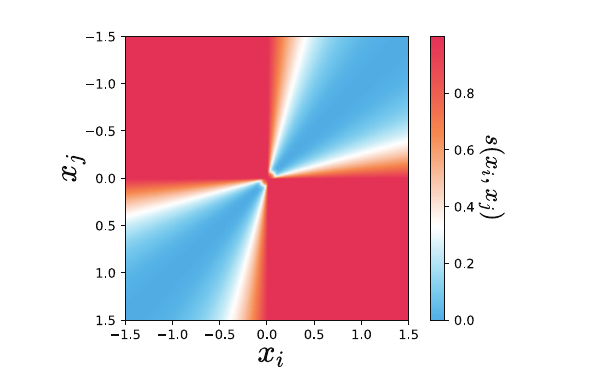}
\caption{Reflection symmetry ratio $s(i,j)$ given two gene expression values: $x_i$ and $x_j.$ For symmetric values, the ratio is 1 and collapses rapidly to 0 when the two values have the same sign.}
\label{fig:fig5_5}
\end{figure}

To apply the idea of symmetry to these data, we normalized the expression of each gene. Specifically, negative (positive) normalized quantities represented a lower (higher) expression level than the average expression found for a given gene in all the cells tested. Genes that are highly expressed in more than 70\% of the cell are removed as these are typically housekeeping genes unrelated to the process of interest. We also remove genes that are not expressed in at least 25\% of the cells since it becomes too hard to find two cells having the same gene expressed. Genes with zero expression are changed to NaN. Finally, genes with a decreasing expression distribution, i.e. where most expression values are close to 0, are removed because of their lack of suitability for defining a symmetry axis.  We then defined a parity metric to determine whether two cells had a symmetric expression at the gene level. This metric quantifies how opposite the same gene expressions are for two different cells.  For each gene, the maximum of the distribution of each gene's expression over all cells is defined as the zero expression level. Redefining the zero expression allows us to think of the expression of each gene as being more (positive value) or less (negative value) than the most common quantity found in the cells. The maximum of the distribution was preferred over the average since the distributions have highly zero-inflated counts due to the experimental technique used. The expression value for each gene is then scaled by the standard deviation. Let $x$ be a gene, $i,j$ be a pair of cells, and $s(i,j)$ be the reflection symmetry ratio. We ask for the function $s(i,j)$ to be 1 for a perfectly symmetric expression, i.e. $x_i=-x_j$. We also desire that the function will output a 0 if $\left| x_i \right|\approx 0$ and $\left| x_j\right| \approx 0$ even if $x_i=-x_j$. This condition allows us to avoid attributing some symmetry properties to noise by requiring a difference of at least 20\% of the standard deviation between the two expressions $x_i$ and $x_j$. Finally, we want the function to output a 0 when $x_i$ and $x_j$ are of the same sign. One possible function with these properties is given by
\begin{align*}
s(i,j)= 1- \max \left\lbrace \left| \frac{x_i + x_j}{ \left|x_i \right| +  \left|x_j \right| + \epsilon }\right|^{h_1} , \frac{1 - \tanh{ h_2\left( r_{x,ij} - r \right) }}{2}  \right\rbrace ,
\end{align*} 
\noindent where $h_1, h_2$ are parameters that select the smoothness of the transition between the values of 0 and 1, and $r_{x,ij} = \sqrt{x_i^2 + x_j^2}$. The parameters used for the indicator are $h_1 = 1.7$, $r = 0.1$, $h_2 = 25$, and $\epsilon = 1\times 10^{-7}$. Figure \ref{fig:fig5_5} shows the reflection symmetry ratio for these parameters. 

We observed that the distribution of the parity metric spanned from 0 (non-symmetric pairs) to 0.4 (symmetric pairs). The distribution was also bi-modal, suggesting the existence of two pair types (see Figure \ref{fig:fig5}.B). The expectation value of the parity metric for two normally distributed gene expression measurements is 0.343; by contrast, we find that the observed gene expression measurements are skewed toward non-symmetry. With these data and metrics in hand, we studied the symmetrical opposition of the two different states (drug-sensitive or primed). Specifically, we computed the probability that a pair of cells respecting a given criterion were of different states. When considering all pairs independently from their reflection symmetry, we observed a probability of approximately 35\% for a pair to have different states. Therefore, we have approximately a one-in-three chance of picking a drug-sensitive and a primed cell when picking a pair of cells at random in our data set. We expected pairs with a high (low) reflection symmetry ratio to have a higher (lower) probability of residing in different paths (drug-sensitive vs. primed). Our expectation was confirmed: pairs with a low reflection symmetry ratio had a probability of around 25\% to be of different states versus 80\% for the high reflection symmetry ratio pairs (see Figure \ref{fig:fig5}.C). Broadly, these data demonstrate that reflection symmetry correlates with drug resistance in cancer and could be used to classify a cell as drug-resistant or drug-responsive. 

The application of symmetry analysis to gene expression inside melanoma scRNA-seq data is a simple example of the potential of considering symmetry in biology. However, the drug-sensitive and drug-resistant paths were not directly observed in the cells where the gene expression was measured. The labeling was inferred by principal component analysis of the cells' gene expression, where the variance was found to be principally located on one axis. This axis was later found to correlate highly with sensitivity and drug resistance. Additional studies could be conducted on data sets that are not linearly separable using one axis and on data expressing more than a binary category. 

\clearpage
\newpage
\bibliography{bib}

\end{document}